\documentclass[useAMS,usenatbib]{mn2e}
\usepackage{amsmath} 
\usepackage{amssymb} 
\usepackage{verbatim}
\usepackage{graphicx}
\usepackage[percent]{overpic}
\usepackage{subfigure}
\usepackage[percent]{overpic}
\usepackage{esint}
\usepackage{url}
\usepackage{mycommands}
\title[G-modes in Layered Semi-Convection]{The Properties of G-modes in Layered Semi-Convection}

\author[Mikhail A. Belyaev, Eliot Quataert and Jim Fuller]{Mikhail A. Belyaev$^1$\thanks{email: mbelyaev@berkeley.edu},  Eliot Quataert$^1$ and Jim Fuller$^{2,3}$
\\$^1$Astronomy Department and Theoretical Astrophysics Center, University of California Berkeley, Berkeley, CA 94720, USA
\\$^2$TAPIR, Walter Burke Institute for Theoretical Physics, Mailcode 350-17, California Institute of Technology, Pasadena, CA 91125, USA
\\$^3$Kavli Institute for Theoretical Physics, Kohn Hall, University of California, Santa Barbara, CA 93106, USA}

\begin{document}

\maketitle

\begin{abstract}
We study low frequency waves that propagate in a region of layered semi-convection.  Layered semi-convection is predicted to be present in stellar and planetary interiors and can significantly modify the rate of thermal and compositional mixing.   We derive a series of analytical dispersion relations for plane-parallel layered semi-convection in the Boussinesq approximation using a matrix transfer formalism.   We find that like a continuously stratified medium, a semi-convective staircase -- in which small convective regions are separated by sharp density jumps -- supports internal gravity waves (g-modes).  When the wavelength is much longer than the distance between semi-convective steps, these behave nearly like g-modes in a continuously stratified medium.  However, the g-mode period spacing in a semi-convective region is systematically {\em smaller} than in a continuously stratified medium, and it decreases with decreasing mode frequency.    When the g-mode wavelength becomes comparable to the distance between semi-convective steps, the g-mode frequencies deviate significantly from those of a continuously stratified medium (the frequencies are higher). G-modes with vertical wavelengths smaller than the distance between semi-convective steps are evanescent and do not propagate in the staircase. Thus, there is a lower cutoff frequency for a given horizontal wavenumber.   We generalize our results to gravito-inertial waves relevant for rapidly rotating stars and planets.   Finally, we assess the prospects for detecting layered semi-convection using astero/planetary seismology.
\end{abstract}

\section{Introduction}
\label{introduction}
Semi-convection is a fundamental physical process with implications for stellar and planetary evolution and structure. In the astrophysical context, semi-convection occurs when an unstable thermal (entropy) gradient is stabilized against overturning convection by a compositional gradient (e.g.\ of helium) \citep{SchwarzschildHarm,Kato66,Langer83,Noels10}. The requisite physical conditions can occur at the boundary between convective and radiative zones in the cores of massive main-sequence stars (e.g.\ B and O stars) \citep{Langer85,Langer91} and helium-burning horizontal branch stars \citep{Castellani71,Sweigart72}; they can also occur in giant planets for which helium and hydrogen are not fully miscible \citep{Stevenson85,Leconte12,Nettelman15}, and there exists a global compositional gradient due to e.g.\ helium rain out \citep{Salpeter73,Stevenson75,Wilson10}. For stars, the presence of semi-convection has significant ramifications for stellar evolution \citep{Weiss89,Stothers94,Noels14}; for planets, its presence modifies the global heat content (entropy) and metal distribution, which might help explain the inflated radii of hot Jupiters \citep{Chabrier07} and the luminosity of Saturn \citep{Leconte13}.

There has been much recent progress in modeling the microphysics of semi-convection using both analytical theory and simulations \citep{Rosenblum11,Mirouh12,Wood13,Zaussinger13}. Semi-convection arises from a double-diffusive instability \citep{Stern60}, specifically when heat diffuses much more rapidly than composition. Double-diffusive instability can either lead to turbulent diffusion or layered semi-convection, depending on the properties of the medium. In this work, we focus on layered semi-convection \citep{Proctor81,Spruit92,Spruit13,Radko03}, which has a staircase structure for the density (see Fig. \ref{density_schematic}) that is a unique signature of semi-convection, at least in certain regimes. Layered semi-convection is analogous to overturning convection but with eddies occurring on the scale of the separation between individual steps in the staircase. The step size separation, $d$, is typically smaller than the scale height, $H$, with estimates for $d/H$ spanning orders of magnitude in range from $10^{-6} \lesssim d/H \lesssim 1$ \citep{Leconte12,Zaussinger13,Nettelman15}. Semi-convection becomes more efficient at tranporting energy with increasing step-size, because of the increasing distance convective eddies can travel before re-depositing their energy.

Layered semi-convection caused by the salt-finger instability has been observed in thermohaline staircases in the ocean. However, definitive observational signatures of layered semi-convection are lacking in the astrophysical context. Given the high resolution stellar asteroseismology data from the {\it Kepler} satellite \citep{Gilliland10,Christensen12}, as well as {\it Cassini} observations of density waves excited in Saturn's rings by internal modes of Saturn \citep{Marley93,Hedman13,Fuller14,Fuller14a}, it is natural to look for signatures of semi-convection in the oscillation modes of stars and planets.

Our goal in this work is to understand the low frequency oscillation modes in layered semi-convection using a local plane-parallel approximation. These low frequency modes have buoyancy and/or rotation as their primary restoring force. We work under the assumption that a semi-convective staircase exists, without concerning ourselves with the underlying (and fascinating) physics of how it is generated and evolves over time. We find that like a continuous, stably-stratified medium, a semi-convective staircase with density jumps supports g-modes that are the analogs of internal gravity waves. G-modes with wavelengths that are long compared to the step size behave like internal gravity waves, though with small corrections that we quantify and which may be detectable. However, g-modes with wavelengths of order the step size are strongly affected by the discrete nature of the density jumps in the staircase, and their dispersion relation differs from that of internal gravity waves. Moreover, the discrete nature of the steps in the staircase introduces a lower cutoff frequency for g-modes with wavelengths smaller than the step size. This cutoff frequency has no counterpart in a continously stratified medium.

We begin by describing the setup of our model for a semi-convective staircase in \S \ref{setup}. In \S \ref{gmodes}, we perturb around this background state and derive analytical expressions for the dispersion relation of linear g-modes subject to different boundary conditions, under the Boussinesq approximation. In \S \ref{rotationsec}, we study the effect of rotation on the dispersion relation of g-modes. In \S \ref{asteroseismology_sec}, we discuss the potential of detecting layered semi-convection using asteroseismology, and we end with a summary and a discussion of our results in \S \ref{discussion}.

\section{Problem Setup}
\label{setup}

We assume a background state that is in hydrostatic equilibrium and adopt a plane parallel approximation, so we ignore curvature and work on scales that are small compared to the radius of the star/planet. Gravity is constant in our setup and points in the $-\zhat$ direction so that the equation of hydrostatic equilibrium is $dP/dz = -g\rho$.

For the density profile, we assume a semi-convective staircase, which has discrete interfaces separating convective zones. At each interface, the density undergoes a discontinuous jump by a value $\Delta \rho$. Since the fluid is convective in between any two interfaces (within a stair), the Brunt-V{\"a}is{\"a}l{\"a} frequency
\begin{align}
\label{Bruntdef}
N^2 \equiv -\frac{1}{\gamma \rho} \frac{dP}{dz} \frac{d \ln P \rho^{-\gamma}}{dz}
\end{align}
within a stair is vanishingly small compared to the frequencies of interest in the problem, so we set it to zero.

For simplicity, we shall typically assume that the distance between adjacent interfaces, $d$, is constant, though we also discuss the effect of relaxing that assumption. We also assume that $d \ll H$, where $H$ is the characteristic length scale for the background equilibrium quantites to change by order unity ($H$ is of order the scale height). Under these assumptions, the magnitude of the jump in density, $\Delta \rho > 0$, between adjacent stairs is given by
\begin{align}
\label{barN}
\frac{g \Delta \rho}{\rho d} = \bar{N}^2,
\end{align}
where $\bar{N}$ is the \BV frequency given by integrating equation (\ref{Bruntdef}) over a single stair, including the interface. Since $N \approx 0$ within a stair due to convection and each interface is presumed to be infinitely thin, the only contribution to $\bar{N}$ is from the jump in density across an interface. Also, because $d \ll H$ by assumption, we generally have $\Delta \rho/\rho \ll 1$.

\begin{figure}
\centering
\includegraphics[width=.49\textwidth]{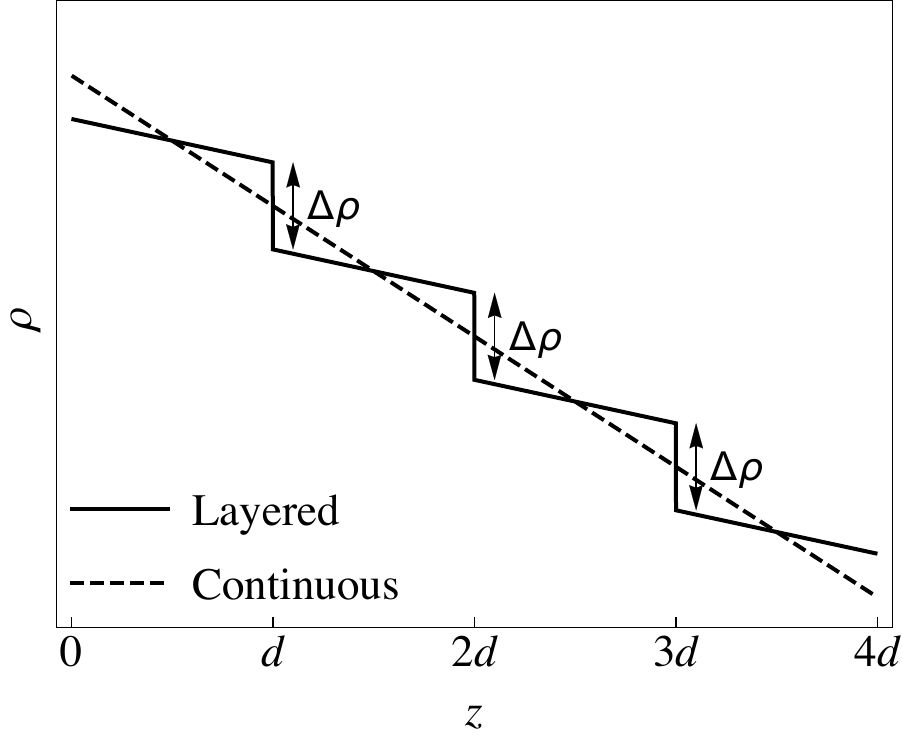}
\caption{Solid curve -- Density profile for layered semi-convection; dashed curve -- Density profile for a stable, continuously stratified medium with \BV frequency $\bar{N}$. Since we shall be working on scales that are small compared to the scale height, we have used straight line segments to approximate both an adiabatic gradient (solid line segments between density jumps) and a continuous density profile.}
\label{density_schematic}
\end{figure}

Figure \ref{density_schematic} shows a schematic of the density profile that we consider for layered semi-convection (solid curve) vs. a continuous density profile with the same value of $\bar{N}$ (dashed line). For the semi-convective profile, the interfaces are spaced in intervals of $d$ and the jump at each interface is $\Delta \rho$. Between interfaces, $\rho(z)$ follows an adiabatic gradient, due to convection between stairs. 

Having described the equilibrium background state, we shall now linearize about it to derive the dispersion relation for g-modes. For simplicity, we ignore any corrections to the dispersion relation resulting from a finite scale height. Thus, we assume the wavelength of the perturbations is much smaller than the scale height ($\lambda/H \ll 1$). This limit together with the low frequency assumption $\omega \ll c_s/\lambda$, where $c_s$ is the sound speed, constitute the Boussinesq approximation.

\section{Dispersion Relation for G-Modes}
\label{gmodes}

We now derive the dispersion relation for the analog of g-modes in a semi-convective staircase under the Boussinesq approximation. We initially neglect rotation, but we return to the effects of rotation in \S \ref{rotationsec}. Assuming an adiabatic equation of state and (without loss of generality) motion in the $x-z$ plane, the linearized Boussinesq equations are
\begin{align}
\label{Boussinesq}
0 &= \frac{\p \dd u}{\p x} + \frac{\p \dd w}{\p z}  \\
\frac{\p \dd u}{\p t} &= -\frac{1}{\rho}\frac{\p \dd P}{\p x} \nn \\
\frac{\p \dd w}{\p t} &= -\frac{1}{\rho}\frac{\p \dd P}{\p z} - g \frac{\dd \rho}{\rho} \nn \\
\frac{\gamma}{\rho} \frac{\p \dd \rho}{\p t} &= \delta w \frac{d \ln P \rho^{-\gamma}}{dz}, \nn
\end{align}
where $\delta u$ and $\delta w$ are the fluid velocities in the $x$ and $z$ directions, respectively.

This set of equations can be simplified further by noting that $d \ln P \rho^{-\gamma}/dz = 0$ within a stair due to convection. Thus, the set of equations (\ref{Boussinesq}) within a stair reduces to
\begin{align}
\label{incompressible}
0 &= \frac{\p \dd u}{\p x} + \frac{\p \dd w}{\p z}  \\
\frac{\p \dd u}{\p t} &= -\frac{1}{\rho}\frac{\p \dd P}{\p x} \nn \\
\frac{\p \dd w}{\p t} &= -\frac{1}{\rho}\frac{\p \dd P}{\p z}, \nn
\end{align}
from which it immediately follows that $\nabla^2 \dd P = 0$

The set of equations (\ref{incompressible}) applies within a stair, but we also need to connect adjacent stairs via boundary conditions at the interfaces. There are two standard boundary conditions in this case: the contact condition, which states that the fluids on either side of the interface must stay in contact, and the force balance condition, which states that the forces must balance across the interface to avoid infinite accelerations. 

Let us assume for specificity that interfaces are located at integer multiples of $d$, and that the $n$-th stair lies between $-nd \le z \le -(n-1)d$. Referencing a perturbation in the $n$-th stair by the subscript $n$, and defining $\xi$ to be the displacement vector of a fluid element from its equilibrium position, the boundary conditions across an interface are
\begin{align}
\label{BCstair}
\xi_{n+1} &= \xi_n \\
\dd P_{n+1} &= \dd P_n + \xi_n g \Delta \rho. \nn
\end{align}
The first boundary condition expresses the requirement that the two fluids stay in contact with each other. The second boundary condition ensures that the momentum flux is continuous across the interface and can be obtained by integrating the $z$-component of the momentum equation across the interface. All quantities in the equations (\ref{BCstair}) are evaluated at the interface located at $z = -nd$, and the boundary conditions connect the linear perturbations in the $n$-th stair to those in the $(n+1)$-th stair.

Next, we would like to express $\xi_n$ in terms of $\delta P_n$. To do so, we Fourier transform equations (\ref{incompressible}) in $x$ and $t$ and assume normal mode perturbations in the form $\delta P_n \propto e^{i (k_\perp  x - \omega t)}$. The $z$-dependence of the eigenfunction follows directly from the relation $\nabla^2 \delta P = 0$. Thus, the form of the pressure perturbation for plane waves in the $x$-direction can be written as
\begin{align}
\label{perturbation_form_n}
\delta P_n = \left(A_n e^{k_\perp (z+nd)} + B_n e^{-k_\perp (z + nd)} \right)e^{i (k_\perp  x - \omega t)}.
\end{align}
Combining equation (\ref{perturbation_form_n}) with the $z$-component of the momentum equation (\ref{incompressible}), we see that
\begin{align}
\label{xidPstair}
\xi_n = \frac{k_\perp }{\rho \omega^2} \left(A_ne^{k_\perp (z+nd)} - B_ne^{-k_\perp (z+nd)}\right) e^{i (k_\perp  x - \omega t)}.
\end{align}
Here, we have assumed that $\Delta \rho/\rho \ll 1$, in order to be consistent with the Boussinesq approximation, which neglects the vertical density variation, except in the buoyancy -- $g \Delta \rho$.

The boundary conditions (\ref{BCstair}) written in terms of $A_n$ and $B_n$ are
\begin{align}
\label{BC_AB}
A_{n+1}e^{k_\perp d}-B_{n+1}e^{-k_\perp d} &= A_n-B_n  \\ \nn
A_{n+1}e^{k_\perp d} + B_{n+1}e^{-k_\perp d} &= A_n + B_n +\frac{gk_\perp  \Delta \rho}{\rho \omega^2}(A_n-B_n).
\end{align}
In addition to these equations, which connect the values of $A$ and $B$ in adjacent stairs across an interface, we must also provide boundary conditions for $A$ and $B$. Once boundary conditions are specified, it is possible to solve for $\omega$ and derive the dispersion relation. In what follows, we consider two different cases: an infinite staircase and a finite staircase embedded in a convective zone.

\subsection{Infinite Staircase}
\label{disrel_periodic}

We now derive the dispersion relation for an infinite semi-convective staircase. We first consider strictly periodic boundary conditions. In other words, we assume that $A_n = A_{n+m}$ and $B_n = B_{n+m}$, so the eigenfunctions repeat every $m$ stairs. More general solutions are given in \S \ref{disrel_infinite}. For a periodicity of $m$ stairs, we see from equations (\ref{BC_AB}) that we will in general have a system of $2m$ equations to solve. Defining 
\begin{align}
W^2 \equiv gk_\perp  \Delta \rho/\rho \omega^2 = (\bar{N}/\omega)^2 k_\perp d,
\end{align} 
this system of equations can be written in matrix form as
\begin{equation}
\label{dis_rel_det}
\begin{split}
\begin{bmatrix}
\bfP & \bfQ & \bfzero & \hdots & \bfzero \\
\bfzero & \ddots & \ddots & \ddots & \vdots  \\
\vdots & \ddots & \ddots & \ddots & \bfzero\\
\bfzero & \ddots & \ddots & \ddots & \bfQ \\
\bfQ  &  \bfzero  & \hdots & \bfzero  & \bfP 
\end{bmatrix} 
\begin{bmatrix}
\bfv_1 \\
\vdots \\
\bfv_m
\end{bmatrix} &= \bfzero, \ \ \bfv_n \equiv \begin{bmatrix}
A_n \\
B_n \\
\end{bmatrix},
\\ \bfP \equiv \begin{bmatrix}
1 & -1 \\
-(1+W^2) & W^2-1 \\
\end{bmatrix} &, \ \ \bfQ \equiv \begin{bmatrix}
-e^{k_\perp d} & e^{-k_\perp d} \\
e^{k_\perp d} & e^{-k_\perp d} \\
\end{bmatrix}.
\end{split} 
\end{equation}
Here, $[\bfv_1, ... \bfv_m]$ is the vector of length $2m$ composed of the coefficients $A_n$ and $B_n$ and $\bfP$ and $\bfQ$ are $2 \times 2$ submatrices. $\bfP$ is repeated down the main diagonal of the $2m \times 2m$ matrix $m$ times, and $\bfQ$ is repeated down the diagonal above it $m-1$ times. There is a corner element, $\bfQ$, in the lower left hand corner, which enforces periodicity, and all other matrix elements are zero. 

Equation (\ref{dis_rel_det}) is not ideal, because the size of the matrix grows without bound as $m$ increases. However, one can write down the solution of block tridiagonal systems in the form (\ref{dis_rel_det}) by making use of a recurrence relation \citep{Molinari08}. For instance, we can write $\{A_{n+1},B_{n+1}\}$ in terms of $\{A_n,B_n\}$ as 
\begin{equation}
\label{transfer_matrix}
\bfT \begin{bmatrix}
A_n \\
B_n
\end{bmatrix} = \begin{bmatrix}
A_{n+1} \\
B_{n+1}
\end{bmatrix},
\end{equation}
where the transfer matrix $\bfT$ is defined by
\begin{equation}
\bfT \equiv \begin{bmatrix}
 \left(1+\frac{W^2}{2} \right)e^{-k_\perp d} & -\frac{1}{2}W^2e^{-k_\perp d} \\
 \frac{1}{2}W^2e^{k_\perp d}   &  \left(1- \frac{W^2}{2} \right)e^{k_\perp d}
\end{bmatrix}.
\end{equation}
The system of equations (\ref{dis_rel_det}) with periodic boundary conditions can be written succintly using the transfer matrix as
\begin{align}
\label{recurrence_relation}
\bfT^m \begin{bmatrix}
A_n \\
B_n
\end{bmatrix} = \begin{bmatrix}
A_{n} \\
B_{n}
\end{bmatrix}.
\end{align}
In order for equation (\ref{recurrence_relation}) to have a non-trivial solution, we must have
\begin{align}
\label{det_tm}
\det(\bfT^m - \bfI) = 0,
\end{align}
where $\bfI$ is the $2 \times 2$ identity matrix. 

One way to solve equation (\ref{det_tm}) is to compute $\bfT^m-\bf I$ explicitly and take the determinant. Computing $\bfT^m$ can be accomplished by diagonalizing $\bfT$, i.e.\ by writing the transfer matrix in the form $\bfT = \bfU \bfD \bfU^{-1}$, where $\bfD$ is a diagonal matrix. The transfer matrix raised to the power of $m$ is then simply given by $\bfT^m = \bfU \bfD^m \bfU^{-1}$. Alternatively, equation (\ref{det_tm}) can be solved by noting that it implies both of the eigenvalues of the matrix $\bfT^m$ are equal to 1, i.e. $\lambda[\bfT^m] = 1$ for both eigenvalues. This is equivalent to the expression $(\lambda[\bfT])^m = 1$ for the eigenvalues of the transfer matrix $\bfT$. 

Solving equation (\ref{det_tm}), we find after some algebra 
\begin{equation}
\begin{split}
\label{long_dis_rel}
&1 = \left(\cosh
   (k_\perp d) -\frac{W^2}{2} \sinh (k_\perp d) ~ \pm \right. \\
   &\left. \sqrt{\sinh (k_\perp d) \left[\left(\frac{W^4}{4}+1\right) \sinh (k_\perp d)-W^2 \cosh (k_\perp d)\right]}\right)^m   
\end{split}
\end{equation}
We can get rid of the functional dependence on $k_\perp d$ in equation (\ref{long_dis_rel}) by assuming solutions of the form $W^2 = 2\coth(k_\perp d) - 2C \csch(k_\perp d)$, where $C$ is a constant to be solved for. Making this substitution, equation (\ref{long_dis_rel}) simplifies to
\begin{equation}
\label{short_dis_rel}
\left(C \pm \sqrt{C^2-1}\right)^m  = 1.
\end{equation}
Equation (\ref{short_dis_rel}) will be satisfied if $C \pm \sqrt{C^2-1} = e^{2\pi i n /m} $, where $n$ is an integer. Solving this equation for $C$ we have
\begin{align}
\label{Ceq}
C = \cos(2\pi n/m).
\end{align}

Note that since $\cos(x) = \cos(-x)$ there are only $m/2 + 1$ distinct values of $C$, where we round down if $m$ is odd. We should in fact choose $0 \le n \le m/2$, because if $n=m/2$ the eigenfunction repeats every other stair, and higher values of $m$ are ``aliased" to lower values. This is analogous to what happens in signal processing when frequencies higher than half the sampling frequency are aliased to lower frequencies (i.e.\ the Nyquist-Shannon sampling theorem). 

\begin{comment}
Solutions for $C$ for $1\le m \le 6$ are
\begin{align}
C = \left\{
  \begin{array}{lr}
    -2 & : m = 1\\
    \left \{ \pm 2 \right \} & : m = 2\\
    \left \{ -2,1 \right \} & : m = 3\\
    \left\{ 0,\pm 2 \right \} & : m = 4\\
    \left\{ -2,(1\pm\sqrt{5})/2 \right \} & : m = 5\\
    \left\{ \pm1,\pm2 \right \} & : m = 6
  \end{array}
\right.
\end{align}
\end{comment}

Given equation (\ref{Ceq}), the dispersion relation for periodic solutions is
\begin{align}
\label{exact_dispersion_relation}
\omega^2 = \bar{N}^2 \left(\frac{k_\perp d}{2\coth(k_\perp d)-2\cos(2\pi n/m)\csch(k_\perp d)}\right),
\end{align}
where $\bar{N}^2$, defined in equation (\ref{barN}), is the background \BV frequency that would correspond to a smooth density gradient in the absence of semi-convection.

\subsubsection{General Dispersion Relation for the Infinite Staircase}
\label{disrel_infinite}
The solution in equation (\ref{exact_dispersion_relation}) came from requiring that the eigenfunctions repeat every $m$ stairs. This strict periodicity requirement led to a discrete spectrum of modes with a finite number of frequencies. We now relax the assumption of strict periodicity and look for solutions with $[A_{n+m},B_{n+m}] = e^{2 \pi i \psi}[A_n,B_n]$. The dispersion relation in this case is given by the expression
\begin{align}
\label{det_tm_gen}
\det(\bfT^m - e^{2 \pi i \psi} \bfI) = 0,
\end{align}
where in general $\psi$ is a complex constant, $\psi \equiv \psi_r + i \psi_i$ with $\psi_r$ restricted to the range $0 \le \psi_r < 1$ (without loss of generality). Physically, $\psi_r$ represents the phase shift a solution undergoes over $m$ stairs. Strict periodicity requires $\psi = 0$ (equation (\ref{det_tm})), but we can still have purely oscillatory solutions under the less stringent requirement of $\psi_i = 0$. Solutions with $\psi_i \ne 0$ are spatially growing/decaying and are thus analogous to evanescent waves. 

Following the derivation given in \S \ref{disrel_periodic}, and again assuming solutions for $W$ in the form $W^2 = 2\coth(k_\perp d) - 2C \csch(k_\perp d)$, where $C$ is a constant, the dispersion relation (\ref{det_tm_gen}) reduces to
\begin{align}
\label{short_dis_rel_gen}
\left(\frac{C \pm \sqrt{C^2-1}}{e^{2 \pi i \psi/m}}\right)^m = 1,
\end{align}
which is the analog of equation (\ref{short_dis_rel}). Taking $C \equiv \cos(\theta)$, where $\theta$ is in general complex, equation (\ref{short_dis_rel_gen}) reduces to
\begin{align}
\label{shorter_dis_rel_gen}
e^{\pm im\theta - 2\pi i \psi} = 1,
\end{align}
which has solutions
\begin{align}
\label{theta_periodic_gen}
\theta = \frac{2 \pi (n \pm \psi_r)}{m} \pm \frac{2\pi i \psi_i}{m},
\end{align}
where $n$ is an integer.

Considering just oscillatory solutions for the time being, which have $\psi_i = 0$, we see that the spectrum of modes for the infinite staircase is actually continuous, and the discreteness obtained in the previous section is an artifact of imposing a periodicity constraint. Thus, the dispersion relation for the infinite staircase can be written as 
\begin{align}
\label{exact_dispersion_relation_gen}
\omega^2 = \bar{N}^2 \left(\frac{k_\perp d}{2\coth(k_\perp d)-2\cos(k_z d)\csch(k_\perp d)}\right),
\end{align}
where we have substituted $\theta \rightarrow k_z d$. The reason for this substitution will become clear in the next section when we treat $k_z$ as the vertical wavevector for a g-mode propagating through the staircase. For purely oscillatory solutions, we have the requirement that $k_z$ is real.

\subsubsection{Analysis of Infinite Staircase Solution}
\label{comparison_sec}

\begin{figure*}
\centering
\subfigure{\begin{overpic}[width=.505\textwidth]{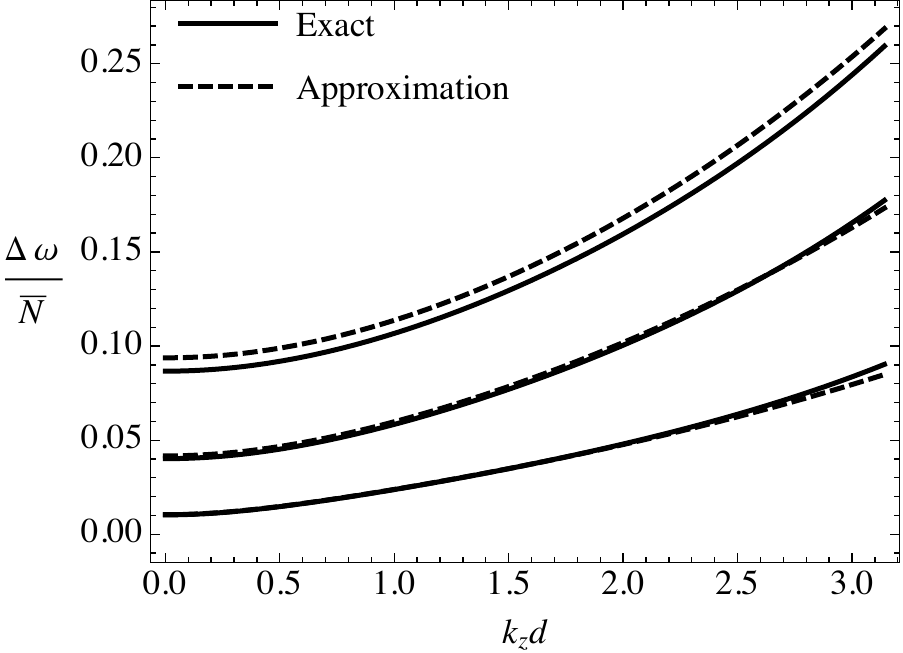}
\put(78.5,31){$k_\perp d = 0.5$}
\put(67,40){$k_\perp d = 1$}
\put(50,48){$k_\perp d = 1.5$}
\end{overpic}}
\subfigure{\begin{overpic}[width=.48\textwidth]{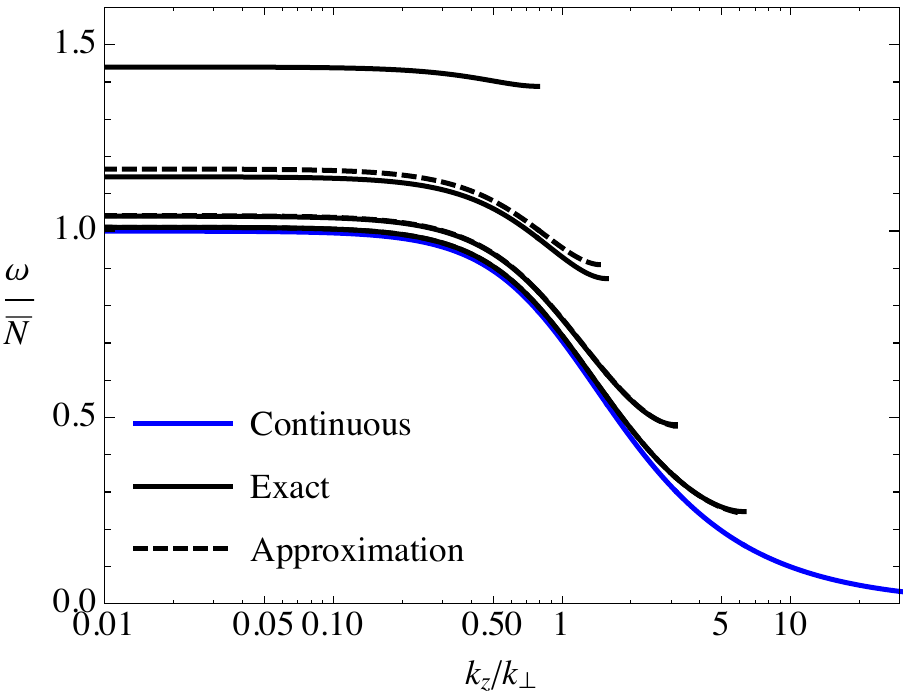}
\put(58,70){$k_\perp d = 4$}
\put(66,50){$k_\perp d = 2$}
\put(74,32){$k_\perp d = 1$}
\put(80,22){$k_\perp d = 0.5$}
\end{overpic}}
\caption{Left: Exact and approximate (series expansion from equation (\ref{deltaomega})) curves for $\Delta \omega/\bar{N}$ as a function of $k_z d$ for $k_\perp d = \{0.5,1,1.5\}$. Right: Plot of of $\omega/\bar{N}$ as a function of $k_z/k_\perp $ for various values of $k_\perp  d$. The continuous limit ($k_\perp d = 0$) given by equation (\ref{omega_0}) is shown by the blue line.}
\label{compare_fig}
\end{figure*}
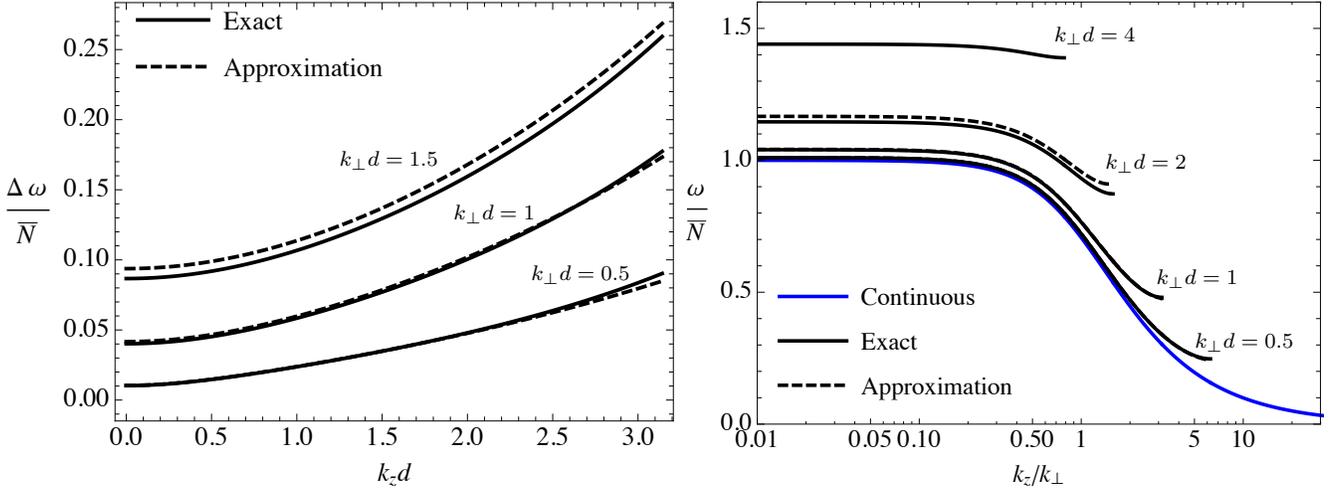

In order to better understand the frequency shifts induced by the presence of a semi-convective staircase compared to a continuous stratification, it is instructive to take the continuous limit of equation (\ref{exact_dispersion_relation_gen}). Taking the limit $\sqrt{k_\perp ^2+k_z^2}d \rightarrow 0$ we can write 
\begin{align}
\label{omega_0}
\omega^2 \approx \bar{N}^2\frac{k_\perp ^2}{k_\perp ^2+k_z^2} + \mathcal{O}\left[ (k_\perp ^2+k_z^2)d^2\right].
\end{align}
The leading order term, which we shall refer to as $\omega_0^2$, gives the dispersion relation for Boussinesq gravity waves in a continuously stratified medium. Thus, the dispersion relation for the semi-convective staircase converges to that of a continuous stratification when the wavelength of a mode is much larger than the distance between stairs.

We can ask how much does the frequency of an eigenmode for the semi-convective staircase deviate relative to that of a Boussinesq gravity wave with the same values of $k_\perp $ and $k_z$? The most straightforward quantity to consider is $\Delta \omega \equiv \omega-\omega_0$, which is the deviation in frequency of the staircase from that of a continuously stratified medium. Expanding $\omega$ in powers of $k_z d$ and $k_\perp  d$, we can approximate the fractional change in frequency as
\begin{align}
\label{deltaomega}
 \frac{\Delta \omega}{\omega_0} \approx \frac{1}{24} \left((k_\perp d)^2 + (k_z d)^2 + \frac{7}{240}(k_z d)^4 \right).
\end{align}
This expression is good to a few percent for values of $k_\perp  d \lesssim 1$ and for any value of $k_z d$. The latter is true because $k_z d \le \pi$, due to aliasing (\S \ref{disrel_periodic}), and higher powers of $k_z d$ are damped by numerical prefactors  for $k_z d \le \pi$. For $k_z d \lesssim 1$, the fourth order term in $k_z d$ can be neglected, but it is needed to provide a good approximation in the range $1 \lesssim k_z d \lesssim \pi$. 

The left panel of Fig \ref{compare_fig} compares the exact value of $\Delta \omega/\bar{N}$ as a function of $k_z d$ for $k_\perp  d = \{0.5,1,1.5\}$ (solid lines) with the series approximation from equation (\ref{deltaomega}) (dashed lines). Even for $k_\perp d = 1$, equation (\ref{deltaomega}) is accurate to within a few percent for $\Delta \omega/\bar{N}$ across the entire range of values for $k_z$. Notice that $\Delta \omega > 0$ always, which can also be seen from equation (\ref{deltaomega}) and implies that the staircase is ``stiffer" than a continuously stratified medium.

The right panel of Fig \ref{compare_fig} compares $\omega/\bar{N}$ as a function of $k_z/k_\perp $ for various values of $k_\perp d$. For a continuous density stratification, $\omega/\bar{N} = \omega_0/\bar{N}$ is a single curve given by equation (\ref{omega_0}).  For a stratified staircase, however, the frequencies are higher than for a continuous density stratification. The black curves corresponding to the staircase dispersion relation in the right panel of Fig \ref{compare_fig} terminate abruptly, because the maximum value of $k_z/k_\perp $ is $\pi/k_\perp d$ as a result of the condition $k_z d \le \pi$.

From the right panel of Fig \ref{compare_fig}, it is clear that the deviation from the continuous case increases with $k_z d$, for a given $k_\perp d$, which is predicted by equation (\ref{deltaomega}). Also, there is a minimum value of $\omega$, which occurs at $k_z d = \pi$. For a given $k_\perp d$, this cutoff frequency is given by
\begin{align}
\label{cutoff_freq}
\frac{\omega_c^2}{\bar{N}^2} = \frac{k_\perp d \sinh(k_\perp d)}{2(\cosh(k_\perp d)+1)}.
\end{align}
In the limit $k_\perp d \ll 1$, the cutoff frequency is $\omega_c \approx \bar{N} k_\perp d/2$. In a continuously stratified medium, this would correspond to the frequency of an internal gravity wave which has a vertical wavelength of order the separation between stairs and a horizontal wavevector equal to $k_\perp $.

Although oscillatory solutions below the cutoff frequency, $\omega_c$, are not allowed, it is possible to have evanescent solutions. These correspond to a complex vertical wavevector, $k_z$. For instance, if we take $k_z d = \pi + i \alpha d $, where $-\infty < \alpha < \infty$, equation (\ref{exact_dispersion_relation_gen}) becomes
\begin{align}
\omega^2 = \bar{N}^2 \left(\frac{k_\perp d}{2\coth(k_\perp d)+2\cosh(\alpha d)\csch(k_\perp d)}\right).
\end{align}
Thus, we can have any value of the frequency in the range $0 < \omega < \omega_c$, by choosing the magnitude of $\alpha$. The sign of $\alpha$ doesn't matter for the dispersion relation, but it does affect whether the waves grow or damp with increasing $z$. From equations (\ref{det_tm_gen}) and (\ref{theta_periodic_gen}), we see that the exponential growth/damping rate per unit length for an evanescent wave is simply $1 / \alpha d$. The interpretation of these evanescent solutions with complex $k_z$ is that if a wave with frequency $\omega < \omega_c$ is incident on a semi-convective region it will be reflected.

Another point worth mentioning about the right panel of Fig \ref{compare_fig} is that the curves for $\omega/\bar{N}$ flatten out for $k_\perp d \gtrsim 1$ and become nearly constant as a function of $k_z$. Indeed, in the limit of $k_\perp  d \gg 1$, the exact dispersion relation (\ref{exact_dispersion_relation}) is independent of $k_z$
\begin{align}
 \omega^2 \approx \frac{\bar{N}^2 k_\perp  d}{2} = \frac{gk_\perp  \Delta{\rho}}{2 \rho}.
\end{align}
This is the standard dispersion relation for an interfacial gravity wave in the Boussinesq approximation ($\Delta \rho/\rho \ll 1$) in an infinitely extended fluid. This makes sense, because the limit $k_\perp d \gg 1$ corresponds to the case when the wavelength is much shorter than the distance between stairs, so the wave dynamics is set by a single interface.

\subsection{Finite Staircase Embedded in a Convective Medium}
\label{disrel_finite}
We now derive the dispersion relation for a semi-convective staircase of finite vertical extent with $m$ interfaces, embedded in a convective medium. We again need to solve a system of equations similar to (\ref{dis_rel_det}), but with boundary conditions $A_0 = 0$ at the top of the staircase and $B_m = 0$ at the bottom. These boundary conditions come from requiring the perturbations to decay as $z \rightarrow \pm \infty$, since the convective medium in which the staircase is embedded does not support g-modes.

Using the transfer matrix $\bfT$ defined in equation (\ref{transfer_matrix}) and applying boundary conditions, we can write
\begin{align}
\label{finite_staircase_1}
\bfT^m \begin{bmatrix}
0 \\
B_0
\end{bmatrix} = \begin{bmatrix}
A_m \\
0
\end{bmatrix}.
\end{align} 
In order for equation (\ref{finite_staircase_1}) to be true in general, the lower right hand corner of the $2 \times 2$ matrix $\bfT^m$ must equal zero. Thus, the dispersion relation becomes
\begin{align}
(\bfT^m)_{11} = 0,
\end{align}
where our row and column indices take the values of $0$ and $1$. Computing $\bfT^m$ explicitly using diagonalization and making the substitution 
\begin{align}
\label{W2_substitution}
W^2 = 2\coth(k_\perp d) - 2 \cos(\theta) \csch(k_\perp d),
\end{align}
the dispersion relation can be written as
\begin{comment}
\begin{align}
\label{long_zero_bdys}
&\left(2 e^{k_\perp d} \sqrt{\sinh (k_\perp d) \left(\left(W^4+4\right) \sinh (k_\perp d)-4 W^2 \cosh
   (k_\perp d)\right)}+e^{2 k_\perp d} (W^2-2)+W^2+2\right) \times \\ \nn
   &\left(-\sqrt{\sinh (k_\perp d) \left(\left(W^4+4\right) \sinh (k_\perp d)-4 W^2 \cosh (k_\perp d)\right)}-W^2 \sinh (k_\perp d)+2
   \cosh (k_\perp d)\right)^m+ \\ \nn
   &\left(2 e^{k_\perp d} \sqrt{\sinh (k_\perp d) \left(\left(W^4+4\right) \sinh
   (k_\perp d)-4 W^2 \cosh (k_\perp d)\right)}-e^{2 k_\perp d} (W^2-2)-W^2-2\right) \times \\ \nn
   & \left(\sqrt{\sinh(k_\perp d) \left(\left(W^4+4\right) \sinh (k_\perp d)-4 W^2 \cosh (k_\perp d)\right)}-W^2 \sinh
   (k_\perp d)+2 \cosh (k_\perp d)\right)^m = 0.
\end{align}
We can make further progress by using the substitution 

in which case equation (\ref{long_zero_bdys}) simplifies to
\end{comment}
\begin{equation}
\begin{split}
\label{simpler_zero_bdy}
&e^{im\theta} \left(\cos(\theta)  \coth (k_\perp d) - \csch(k_\perp d) + i\sin(\theta)\right) = \\
&e^{-im\theta}  \left(\cos(\theta)  \coth
   (k_\perp d) - \csch(k_\perp d) - i\sin(\theta)\right).
\end{split}
\end{equation}

Equation (\ref{simpler_zero_bdy}) is of the form
\begin{align}
\label{fancy_form}
e^{i m \theta} Be^{i \phi} - e^{-i m \theta}Be^{-i \phi} = 0,
\end{align}
so we can simplify it to read
\begin{align}
\label{theta_implicit}
\sin(m\theta + \phi) = 0,
\end{align}
where $0 \le \phi < \pi$ is implicitly defined through equations (\ref{simpler_zero_bdy}) and (\ref{fancy_form}). We can convert equation (\ref{theta_implicit}) into an explicit expression for $\theta$ by using the trigonometric angle addition formula. Assuming $\phi \ne 0$ and $\theta \ne 0$, we can write
\begin{align}
\label{theta_explicit}
\cot(m\theta) + \cot \theta \coth(k_\perp d) - \csc \theta \csch(k_\perp d) = 0,
\end{align}
where we have used the relation $\cot(\phi) = \cot \theta \coth(k_\perp d) - \csc \theta \csch(k_\perp d)$.

We can turn equation (\ref{theta_explicit}) into a polynomial expression for $\cos(\theta)$ by using the relations 
\begin{align}
\cos(m\theta) &= T_m(\cos \theta) \\
\sin(m\theta) &=  U_{m-1}(\cos \theta) \sin \theta \nn,
\end{align}
where $T_m$ and $U_m$ are Chebyshev polynomials of the first and second kind, respectively. Using these identities, equation (\ref{theta_explicit}) can be written as a polynomial in $\cos(\theta)$
\begin{multline}
\label{polynomial_roots}
T_m(\cos \theta) ~ + \\
(\cos \theta \coth(k_\perp d) - \csch(k_\perp d))U_{m-1}(\cos \theta) = 0.
\end{multline} 

The dispersion relation is given by equation (\ref{W2_substitution}), where $\cos(\theta)$ takes the values of the roots of the polynomial equation (\ref{polynomial_roots}). Since equation (\ref{polynomial_roots}) is an $m$-th degree polynomial, there will in general be $m$ branches of the dispersion relation for an embedded staircase with $m$ interfaces.

An alternative to equation (\ref{polynomial_roots}) is to use equation (\ref{theta_implicit}) to solve for $\theta$ directly. In particular, we can write
\begin{align}
\label{theta_direct}
\theta  = \frac{\pi n}{m} - \frac{\phi}{m}.
\end{align}
Taking the limits of $m \gg 1$ (large number of steps) and $n \gg 1$ (small vertical wavelength relative to size of staircase), we see that $\theta = \pi n/m + \mathcal{O}(m^{-1})$, since $0 \le \phi < \pi$. In this limit, $\cos\theta \approx \cos(\pi n /m)$, and the dispersion relation for a finite staircase with $m$ interfaces converges to the dispersion relation for an infinitely extended staircase (equation (\ref{exact_dispersion_relation})), assuming a periodicity of $2m$ stairs. 

The convergence to the infinite staircase is not surprising, because in the limit $n \gg 1$ many vertical wavelengths fit between the upper and lower boundaries, and their frequencies are only weakly influenced by the boundary conditions. However, one might wonder why $\cos\theta \rightarrow \cos(\pi n /m)$ rather than $\cos(2 \pi n/m)$ (i.e.\ why the finite staircase with $m \gg 1$ and $n \gg 1$ is equivalent to a staircase of periodicity $2m$ rather than $m$). The reason is that the finite staircase allows ``antiperiodic" solutions as well as strictly periodic ones. An antiperiodic solution of $m$ stairs obeys the equation
\begin{align}
\bfT^m \begin{bmatrix}
A_n \\
B_n
\end{bmatrix} = -\begin{bmatrix}
A_{n} \\
B_{n}
\end{bmatrix}.
\end{align}
The analog of equation (\ref{recurrence_relation}) that permits both periodic and antiperiodic solutions is
\begin{align}
\det(\bfT^m \pm \bfI) = 0.
\end{align}
The solution to this equation is 
\begin{align}
\label{W2_periodic_antiperiodic}
W^2 = 2\coth(k_\perp d) - 2 \cos(\pi n/m) \csch(k_\perp d),
\end{align}
which contains a factor of $\cos(\pi n/m)$ rather than $\cos(2 \pi n/m)$.

\begin{comment}
Taking now the limit $k_\perp d \ll 1$ (with no assumptions on $m$), we see from equation (\ref{simpler_zero_bdy}) that to leading order in $k_\perp d$
\begin{align}
\phi  \approx \frac{\sin \theta}{\cos \theta - 1} k_\perp d.
\end{align}
We can use this relation together with equation (\ref{theta_direct}) to write down the $\cos \theta$ term in the dispersion relation to leading order in $k_\perp d$:
\begin{align}
\label{costheta_leading_order}
\cos(\theta) \approx \cos \left(\frac{\pi n}{m}\right) + \left(\frac{\sin^2(\pi n/m)}{\cos(\pi n/m)-1}\right)k_\perp d.
\end{align}
Equation (\ref{costheta_leading_order}) can be used to study frequency shifts in the $k_\perp d \ll 1$ limit for the embedded staircase of finite extent to leading order in $k_\perp d$.
\end{comment}

\subsection{Nonuniform Step Sizes}
Up to now we have assumed that the step-size, $d$, is a constant. We now relax that assumption and ask what happens when the step size is nonuniform? Assuming periodic solutions, we find that $\omega$ is an {\it even} function of $k_\perp  \bar{d}$, where $\bar{d}$ is a characteristic value of the step size derived by an appropriate average over the the step size distribution. Thus, for $k_\perp  \bar{d} \rightarrow 0$, the frequency deviation from the continuous case must be at least second order in $k_\perp \bar{d}$, which we showed explicitly for the case of equal step sizes (equation (\ref{deltaomega})). 

To treat the problem of unequal step-sizes, we can still use the matrix transfer formalism. In particular, if the step size of the $n$-th stair is $d_n$, equation (\ref{det_tm}) can be written as
\begin{align}
\label{det_uneven}
\det(\bfT_1\bfT_2...\bfT_m - \bfI) = 0,
\end{align}
where the transfer matrix $\bfT_n$ is the same as the transfer matrix $\bfT$ from equation (\ref{transfer_matrix}), but with $d$ replaced by $d_n$. It is also useful to define $d_n \equiv a_n \bar{d}$ where the $a_n$ are numerical coefficients. Thus, the $n$-th transfer matrix can be written as
\begin{align}
\bfT_n \equiv \begin{bmatrix}
\frac{1}{b_n} \left(1+\frac{W_n^2}{2} \right) & -\frac{W_n^2}{2b_n} \\
\frac{b_n W_n^2}{2}   & b_n \left(1- \frac{W_n^2}{2} \right)
\end{bmatrix},
\end{align}
where $b_n = \exp(a_n k_\perp  \bar{d})$ and $W_n^2 = a_n (\bar{N}/\omega)^2 k_\perp  \bar{d}$.

Equation (\ref{det_uneven}) is an implicit equation for $\omega(k_\perp \bar{d},a_1,...,a_m)$. To show that $\omega$ is an even function of $k_\perp \bar{d}$, with the $a_n$ coefficients fixed, we need to demonstrate that $\omega(k_\perp \bar{d}) = \omega(-k_\perp \bar{d})$. Since $\omega(k_\perp \bar{d})$ satisfies equation (\ref{det_uneven}), it is clear that $\omega(-k_\perp \bar{d})$ also satisfies equation (\ref{det_uneven}), but with $b_n \rightarrow 1/b_n$ and $W^2 \rightarrow -W^2$. Thus, the analog of equation (\ref{det_uneven}) for $\omega(-k_\perp \bar{d},a_1,...,a_m)$ is
\begin{align}
\label{det_uneven_star}
\det(\bfT^*_1\bfT^*_2...\bfT^*_m - \bfI) = 0,
\end{align}
where
\begin{align}
\bfT^*_n \equiv \begin{bmatrix}
b_n \left(1-\frac{W_n^2}{2} \right) &  \frac{b_n W_n^2}{2} \\
-\frac{W_n^2}{2b_n}   & \frac{1}{b_n} \left(1+ \frac{W_n^2}{2} \right)
\end{bmatrix}.
\end{align}

We see that $\bfT^*_n$ is the same as $\bfT_n$ but with the indices on both the row and columns flipped (i.e.\ $T_{rc} = T^*_{1-c,1-r}$). From the definition of the $2\times2$ determinant, it is clear that equations (\ref{det_uneven}) and (\ref{det_uneven_star}) are equivalent. Consequently, $\omega(-k_\perp \bar{d})$ also satisfies equation (\ref{det_uneven}), which concludes the proof that $\omega$ must be an even function of $k_\perp \bar{d}$. 

For a staircase with a finite number of steps, one may also wonder whether going to a solution with nonuniform step size, but the same number of steps, changes the number of solutions for $\omega$ that satisfy the dispersion relation. Regardless of the boundary conditions, the maximum number of distinct solutions for $\omega^2$ (or alternatively for $\omega$ with the assumption $\omega \ge 0$) is equal to $m$, the number of steps. This is because multiplication of an input vector by a set of $m$ transfer matrices effectively generates an $m$-th degree polynomial for $\omega^2$, since each transfer matrix is linear in $1/\omega^2$. The dispersion relation for a finite staircase embedded in a convective medium has $m$ distinct roots (e.g. equation (\ref{theta_implicit})), which means it is complete. Going to non-uniform step sizes would shift the positions of these roots, but the number of solutions would remain the same. This conclusion explicitly demonstrates that non-uniform step sizes cannot produce ``splitting" of mode frequencies.

\section{Rotation Effects}
\label{rotationsec}
We now address how to include the effect of a constant rotation rate in the dispersion relation. We assume that the rotation rate is far below breakup and so the centrifugal force can be ignored.   In this limit, the Coriolis force is the key new restoring force and introduces inertial waves into the problem.  The linearized fluid equations inside a stair in the corotating frame are
\begin{align}
\label{rotation}
\bfnabla \cdot \dd \bfv &= 0 \\
\frac{\partial \dd \bfv}{\partial t} &= -\frac{\bfnabla \dd P}{\rho} -2 \bfOmega \times \dd \bfv\nn.
\end{align}
The two major differences with respect to equations (\ref{incompressible}) are the inclusion of the Coriolis force and that we can no longer assume the motion is in a 2D plane. The latter is true if $\bfOmega$ is misaligned with the direction of density stratification. 

Fourier transforming in time (i.e.\ $\p/\p t \rightarrow -i \omega$) and changing variables to the fluid displacement vector, $\bfxi$, defined via
\begin{align}
\dd \bfv = \frac{\p \bfxi}{\p t} = -i \omega \bfxi, 
\end{align}
equations (\ref{rotation}) can be written as
\begin{align}
\label{rot_continuity}
\bfnabla \cdot \bfxi &= 0 \\
\label{rot_momentum}
-\omega^2 \bfxi &= -\frac{\bfnabla \dd P}{\rho} + 2 i \omega (\bfOmega \times \bfxi).
\end{align}
Taking the curl of equation (\ref{rot_momentum}) we have
\begin{align}
-\omega^2 (\bfnabla \times \bfxi) &= -2 i \omega(\bfOmega \cdot \bfnabla) \bfxi,
\label{rot_momentum_2}
\end{align}
and taking the curl again we have
\begin{align}
\label{rot_momentum_3}
\omega^2 \bfnabla^2 \bfxi &= -2 i \omega (\bfOmega \cdot \bfnabla) (\bfnabla \times \bfxi).
\end{align}
Substituting equation (\ref{rot_momentum_2}) into (\ref{rot_momentum_3}), we finally arrive at
\begin{align}
\omega^2 \bfnabla^2 \bfxi &= 4 (\bfOmega \cdot \bfnabla)  (\bfOmega \cdot \bfnabla) \bfxi
\end{align}

Assuming solutions in the form
\begin{align}
\label{rot_func_form}
\bfxi = f(z)\exp[i (\bfk_\perp  \cdot \bfx - \omega t)], 
\end{align}
where $\bfk_\perp $ is the wavevector in the $x-y$ plane (i.e. the plane perpendicular to the stratification) and substituting equation (\ref{rot_func_form}) into equation (\ref{rot_momentum_3}), we have the following second order ODE for $f(z)$
\begin{multline}
\omega^2 \left( -k_\perp ^2 + \frac{d^2}{dz^2} \right) f(z) = \\
4 \left(ik_\perp  \Omega_\perp  + \Omega_z \frac{d}{dz} \right) \left(ik_\perp  \Omega_\perp  + \Omega_z \frac{d}{dz} \right) f(z),
\end{multline}
where $\Omega_\perp  \equiv \bfOmega \cdot \bfk_\perp /k_\perp $.

The solutions to this ODE have the form $f(z) \propto \exp(k_z z)$, where $k_z$ is defined as
\begin{align}
\label{kzspin}
k_z \equiv k_\perp  \left[ \frac{4i \Omega_\perp  \Omega_z \pm \omega \sqrt{\omega^2 - 4\left(\Omega_\perp ^2 + \Omega_z^2\right)}}{\omega^2 - 4\Omega_z^2} \right].
\end{align}
In what follows, we restrict our attention to $\bfOmega$ parallel to $\zhat$ and $\bfOmega$ perpendicular to $\zhat$ and focus on the infinite staircase.

\subsection{Rotation Axis Parallel to $z$}
\begin{figure*}
\centering
\subfigure{\includegraphics[width=.5\textwidth]{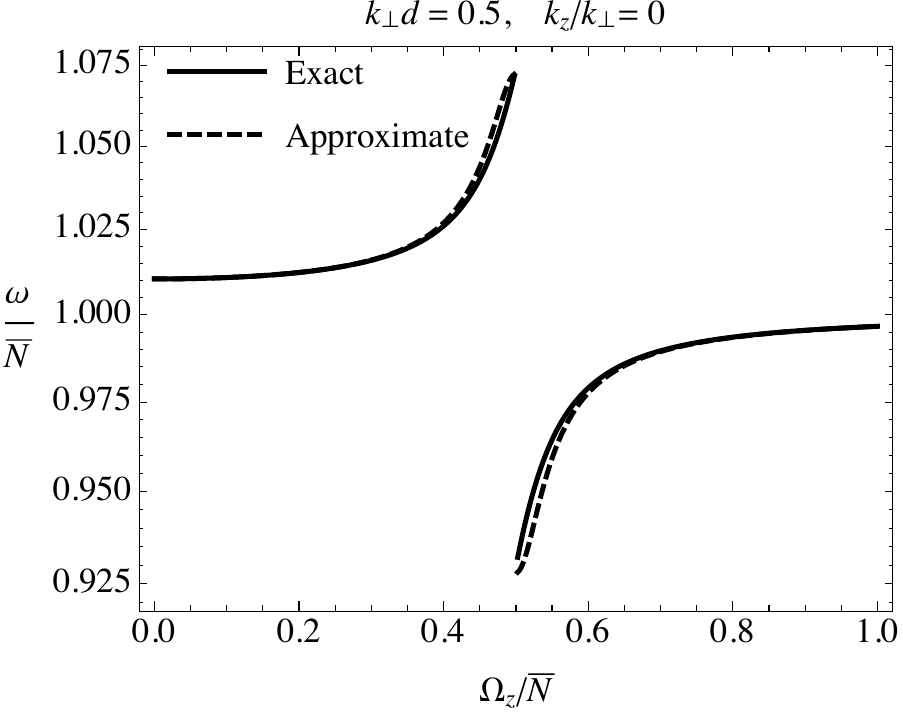}}
\subfigure{\includegraphics[width=.4825\textwidth]{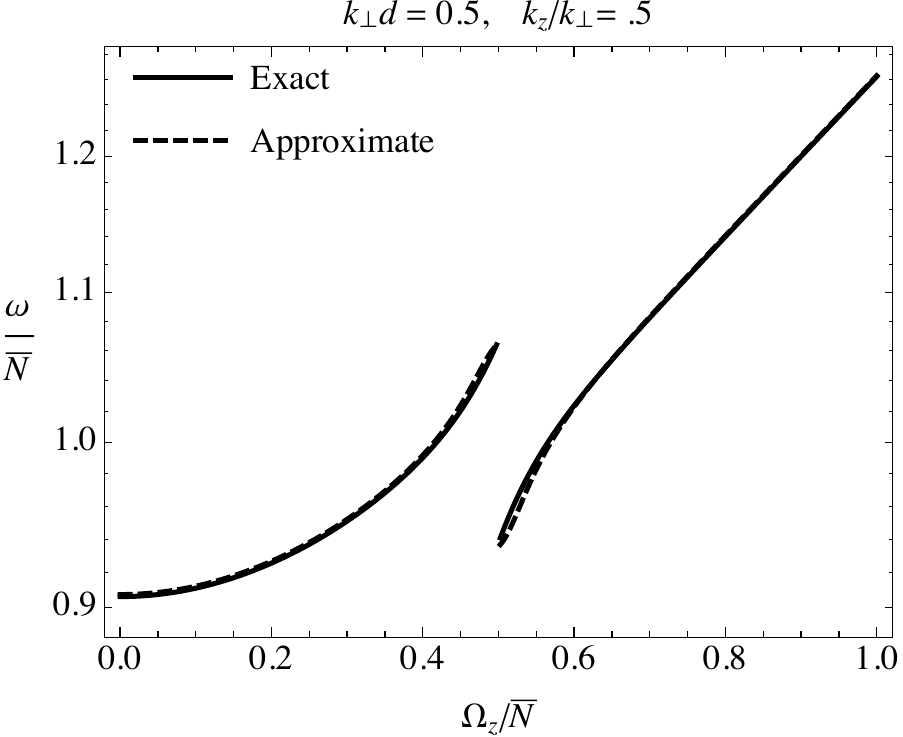}}
\caption{Comparison of exact and approximate solutions for $\omega/\bar{N}$ using the exact dispersion relation (\ref{exact_disrel_aligned}) and an inverse RMS weighting of approximations (\ref{deltaomega_aligned}) and (\ref{dw_special_aligned}). The left panel shows $k_z/k_\perp  = 0$, and the right panel shows $k_z/k_\perp =0.5$.}
\label{domega_aligned}
\end{figure*}

As we show in Appendix \ref{aligned_app}, the boundary conditions between stairs, equations (\ref{BCstair}) and (\ref{xidPstair}), are unchanged for a rotation axis aligned with $z$ if we replace $k_\perp $ with $\bar{k}$:
\begin{align}
\label{k_aligned}
\bar{k} \equiv k_\perp  \left( \frac{\omega}{\sqrt{\omega^2 - 4\Omega_z^2}} \right).
\end{align}
Thus, for an infinite staircase, the dispersion relation can be written as
\begin{align}
\label{exact_disrel_aligned}
\omega^2 = \bar{N}^2 \left(\frac{\bar{k} d}{2\coth(\bar{k} d)-2\cos(k_z d)\csch(\bar{k} d)}\right).
\end{align}

Because equation (\ref{exact_disrel_aligned}) has the same form as equation (\ref{exact_dispersion_relation_gen}), the results of \S \ref{comparison_sec} carry over. For instance, the analog of equation (\ref{omega_0}) in the continuous limit, $\bar{k}d \ll 1$, is
\begin{align}
\omega^2 = \bar{N}^2\frac{\bar{k}^2}{\bar{k}^2+k_z^2} + \mathcal{O}\left[ (\bar{k}d)^2\right].
\end{align}
Substituting for $\bar{k}$ and using the definitions of $k_\perp $, $k_z$, and $\omega_0$ from \S \ref{comparison_sec}, we have 
\begin{align}
\omega_0^2 = \bar{N}^2\frac{k_\perp ^2 \omega_0^2}{k_\perp ^2\omega_0^2+k_z^2(\omega_0^2 - 4\Omega_z^2)}. 
\end{align}
Solving for $\omega_0$ yields
\begin{align}
\label{omega0_aligned}
\omega_0^2 = \frac{k_\perp ^2 \bar{N}^2 + 4 k_z^2 \Omega_z^2}{k_\perp ^2 + k_z^2}.
\end{align}
The exact Boussinesq dispersion relation for plane waves in a continuously stratified medium and an arbitrary orientation of $\bfOmega$ with respect to the direction of stratification is
\begin{align}
\label{omega_exact_rotation}
\omega^2 = \frac{k_\perp ^2 \bar{N}^2 + 4 (\bfk_\perp  \cdot \bfOmega)^2}{k_\perp ^2 + k_z^2}.
\end{align}
Comparing this with equation (\ref{omega0_aligned}), we see that the dispersion relation (\ref{exact_disrel_aligned}) correctly recovers the continuous limit.

In addition to computing $\omega_0$, we can also compute the leading order approximation for the effective wavenumber by plugging equation (\ref{omega0_aligned}) into equation (\ref{k_aligned}): 
\begin{align}
\label{k0_aligned}
\bar{k}_0^2 = k_\perp ^2 \left[\frac{1 + (2\Omega_z/N)^2(k_z/k_\perp )^2}{1-(2\Omega_z/N)^2}\right].
\end{align}
We see that $\bar{k}_0^2 \ge k_\perp ^2$ with equality only if $\Omega_z = 0$ and that $\bar{k}_0$ diverges for $\bar{N}^2 = 4 \Omega_z^2$.

Next, we compute the leading order correction to the dispersion relation in the limit $\bar{k}d \rightarrow 0$. Expanding $\omega$ to second order in $\bar{k}d$ for $\bar{k}d \ll 1$, we have
\begin{align}
\label{deltaomegapure_aligned}
\frac{\omega}{\bar{N}} = \sqrt{\frac{\bar{k}^2}{\bar{k}^2+k_z^2}} + \frac{1}{24} \frac{(\bar{k}_0d)^2}{\omega_0/\bar{N}},
\end{align}
where $\omega_0$ is given by equation (\ref{omega0_aligned}). Note that we cannot replace the first term on the right hand side of equation (\ref{deltaomegapure_aligned}) with $\omega_0$, because $\bar{k}$ contains deviations from $\bar{k_0}$ (equation (\ref{k0_aligned})) at second order in $k_\perp d$ and ultimately we want an expression in terms of $k_\perp d$ not $\bar{k}d$. Solving for $\Delta \omega/\omega_0$ gives
\begin{align}
\label{deltaomega_aligned}
\frac{\Delta \omega}{\omega_0} \approx \frac{1}{24} \left(\frac{(k_\perp  d)^2}{1-(2\Omega_z/\bar{N})^2}\right)\frac{\bar{N}^2}{\omega_0^2}.
\end{align}
Note that for $\bar{N} \ll \Omega_z$, $\omega_0$ is finite but $\Delta \omega \rightarrow 0$. This limit represents purely inertial waves unaffected by the staircase.

The major modification to $\Delta \omega$ in equation (\ref{deltaomega_aligned}) relative to equation (\ref{deltaomega}) is the $1-(2\Omega_z/\bar{N})^2$ term in the denominator, which makes $\Delta \omega/\omega_0$ blow up for $\bar{N}^2 = 4 \Omega_z^2$. However, $\bar{k}_0$ also blows up for $\bar{N}^2 = 4 \Omega_z^2$ (equation (\ref{k0_aligned})), so our expansion in $\bar{k}d \ll 1$ is not valid. To understand what happens when $\bar{N}^2 = 4 \Omega_z^2$, we can set $\omega = \bar{N} + \Delta \omega$, since $\omega_0 = \bar{N}$ is an exact solution to equation (\ref{omega0_aligned}) in this case. Thus, $\Delta \omega$ represents the difference in frequency between the staircase and a continuously stratified medium. From equation (\ref{exact_disrel_aligned}), one finds that for $\bar{k}d \ll 1$
\begin{equation}
\label{dw_special_aligned}
\begin{split}
\frac{\Delta \omega}{\bar{N}} \approx \pm \frac{1}{4 \sqrt{3}} \left(\frac{k_\perp d}{\sqrt{1+(k_z/k_\perp )^2}}\right) \text{ if  } \omega_0^2=\bar{N}^2 = 4 \Omega_z^2,
\end{split}
\end{equation} 
and in particular, there is no divergence in $\Delta \omega$. Additionally, there is no divergence in $\bar{k}$ for $\bar{N}^2 = 4\Omega_z^2$, and $\bar{k}$ can be estimated by plugging $\omega = \bar{N}+\Delta \omega$ with $\Delta \omega$ given by equation (\ref{dw_special_aligned}) into equation (\ref{k_aligned}).

Fig \ref{domega_aligned} shows the exact value of $\omega$ and the approximation $\omega = \omega_0 + \Delta \omega$ for the staircase as a function of $\Omega_z/\bar{N}$. For this figure, $\Delta \omega$ is computed using an inverse root mean square weighting that combines approximations (\ref{deltaomega_aligned}) and (\ref{dw_special_aligned}). 

\subsection{Rotation Axis Perpendicular to $z$}
In the case when the spin axis is perpendicular to $z$ (i.e.\ in the $x - y$ plane), we can assume without loss of generality that $\bfOmega = \Omega_x \xhat$. The jump conditions at an interface are now different from those given in equations (\ref{BCstair}) and (\ref{xidPstair}), and we derive the jump conditions for $\bfOmega$ perpendicular to $\zhat$ in Appendix \ref{misaligned_app}. Using these jump conditions and defining
\begin{align}
\label{k_misaligned}
\bar{k} \equiv  k_\perp  \left( \frac{\sqrt{\omega^2 - 4\Omega_x^2\left(k_x/k_\perp \right)^2}}{\omega} \right),
\end{align}
the dispersion relation for $\bfOmega = \Omega_x \xhat$ can be written as
\begin{align}
\label{exact_disrel_misaligned}
\omega^2 = \bar{N}^2 \left(\frac{k_\perp ^2}{\bar{k}^2}\right) \left(\frac{\bar{k} d}{2\coth(\bar{k} d)-2\cos(k_z d)\csch(\bar{k} d)}\right).
\end{align}
Equation (\ref{exact_disrel_misaligned}) can be rearranged to read
\begin{multline}
\label{exact_disrel_misaligned2}
\omega^2 - 4 \Omega_x^2 \left(\frac{k_x^2}{k_\perp ^2}\right) = \\ \bar{N}^2 \left(\frac{\bar{k} d}{2\coth(\bar{k} d)-2\cos(k_z d)\csch(\bar{k} d)}\right).
\end{multline}

We can check that in the continuous limit, expression (\ref{exact_disrel_misaligned2}) reproduces the dispersion relation for a continuous density stratification with a rotation axis perpendicular to $z$. Taking $\bar{k}d \rightarrow 0$, we have
\begin{align}
\label{continuous_limit_misaligned}
\omega^2 = 4 \Omega_x^2 \left(\frac{k_x^2}{k_\perp ^2}\right) + \bar{N}^2\frac{\bar{k}^2}{\bar{k}^2+k_z^2} + \mathcal{O}\left[ (\bar{k}d)^2\right].
\end{align}
Expanding $\bar{k}$ according to equation (\ref{k_misaligned}), equation (\ref{continuous_limit_misaligned}) reduces to
\begin{align}
\label{omega0_misaligned}
\omega_0^2 = \frac{\bar{N}^2 k_\perp ^2 + 4 \Omega_x^2 k_x^2}{k_\perp ^2 + k_z^2}.
\end{align}
Upon comparison with equation (\ref{omega_exact_rotation}), it is clear that the dispersion relation (\ref{exact_disrel_misaligned}) correctly recovers the continuous limit.

We can also solve for $\bar{k}$ to leading order
\begin{align}
\label{k0_misaligned}
\bar{k}_0^2 = k_\perp ^2\left(1 - \frac{(2\Omega_x/\bar{N})^2(k_x/k_\perp )^2 \left(1+(k_z/k_\perp )^2\right)}{1+(2\Omega_x/\bar{N})^2(k_x/k_\perp )^2}\right),
\end{align}
from which we see that $\bar{k}_0^2 \le k_\perp ^2$ with strict equality only if $\Omega_x k_x = 0$.

\begin{figure}
\centering
\includegraphics[width=.49\textwidth]{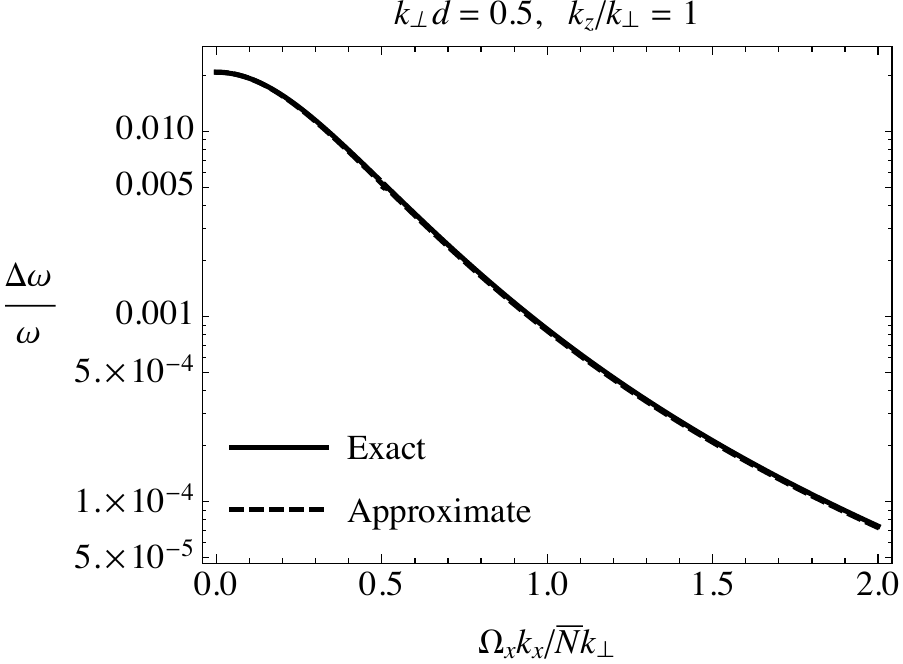}
\caption{Fractional frequency difference, $\Delta \omega/\omega$, between a staircase with rotation perpendicular to stratification and a continuously stratified medium for $k_\perp d = .5$ and $k_z/k_\perp  = 1$. There is excellent agreement between the exact dispersion relation, equation (\ref{exact_disrel_misaligned}), and the approximate solution, equation (\ref{deltaomega_misaligned}).}
\label{domega_misaligned}
\end{figure}

Next, we solve for the leading order corrections to $\omega$ and find that for $\bar{k}d \to 0$
\begin{align}
\label{deltaomega_misaligned}
\frac{\Delta \omega}{\omega_0} &= \frac{1}{24} \left(\frac{(k_\perp ^2 + k_z^2) d^2}{\left[1+(2\Omega_x/\bar{N})^2(k_x/k_\perp )^2 \right]^2}\right).
\end{align}
In contrast to the case when the spin axis is aligned with $z$, there is no divergence of $\Delta \omega/\omega_0$, since $\bar{k}_0$ does not diverge (equation (\ref{k0_misaligned})). Also notice from equations (\ref{deltaomega_misaligned}) and (\ref{deltaomega}) that $\Delta \omega/\omega_0$ is always smaller (for the same values of $k_\perp d$ and $k_z d$) for a spin axis perpendicular to $z$ compared to the case without rotation. Fig \ref{domega_misaligned} shows $\Delta \omega/\omega$ as a function of $\Omega_x k_x/\bar{N} k_\perp $ for $k_z/k_\perp  = 1$ and $k_\perp d = .5$. The approximate and exact solutions are nearly on top of each other.

\section{Application to Asteroseismology}
\label{asteroseismology_sec}

The previous sections show that the presence of layered semi-convection introduces quantifiable wavelength-dependent changes to the dispersion relation for g-modes compared to a continuously stratified medium. However, in order to find an unambiguous signature of semi-convection in asteroseismic (or planetary seismic) data, it is necessary to incorporate our results into a realistic calculation of a stellar/planetary mode oscillation spectrum. A large {\it a priori} uncertainty in such an analysis is the number of density jumps per scale height in layered semi-convection, $H/d$, which can be anywhere from a few to $10^6$ \citep{Leconte12,Zaussinger13,Nettelman15}. 

When $H/d$ is of order unity, one can incorporate the matrix transfer formalism into global seismology calculations directly by computing the jump in the eigenfunction across each step or by modeling the steps as sharp but continuous increases in the density and directly integrating the differential equations across them. However, modeling the jumps individually is likely to become prohibitively expensive as $H/d$ becomes large. In that case, \S \ref{gmodes} and \ref{rotationsec} show that the layered semi-convective medium has a modified dispersion relation compared to a continuously stratified medium.

A unique signature of g-modes in asteroseismology is typically that adjacent global normal modes have a nearly fixed period spacing. However, there are corrections due to e.g.\ a finite scale height and a cavity size that depends on radial wavenumber. Here, we assess the effect of layered semi-convection on the period spacing relation of g-modes compared to a continuously stratified medium, in order to deduce the seismological signatures of layered semi-convection. 

\begin{figure*}
\centering
\subfigure{\includegraphics[width=.48\textwidth]{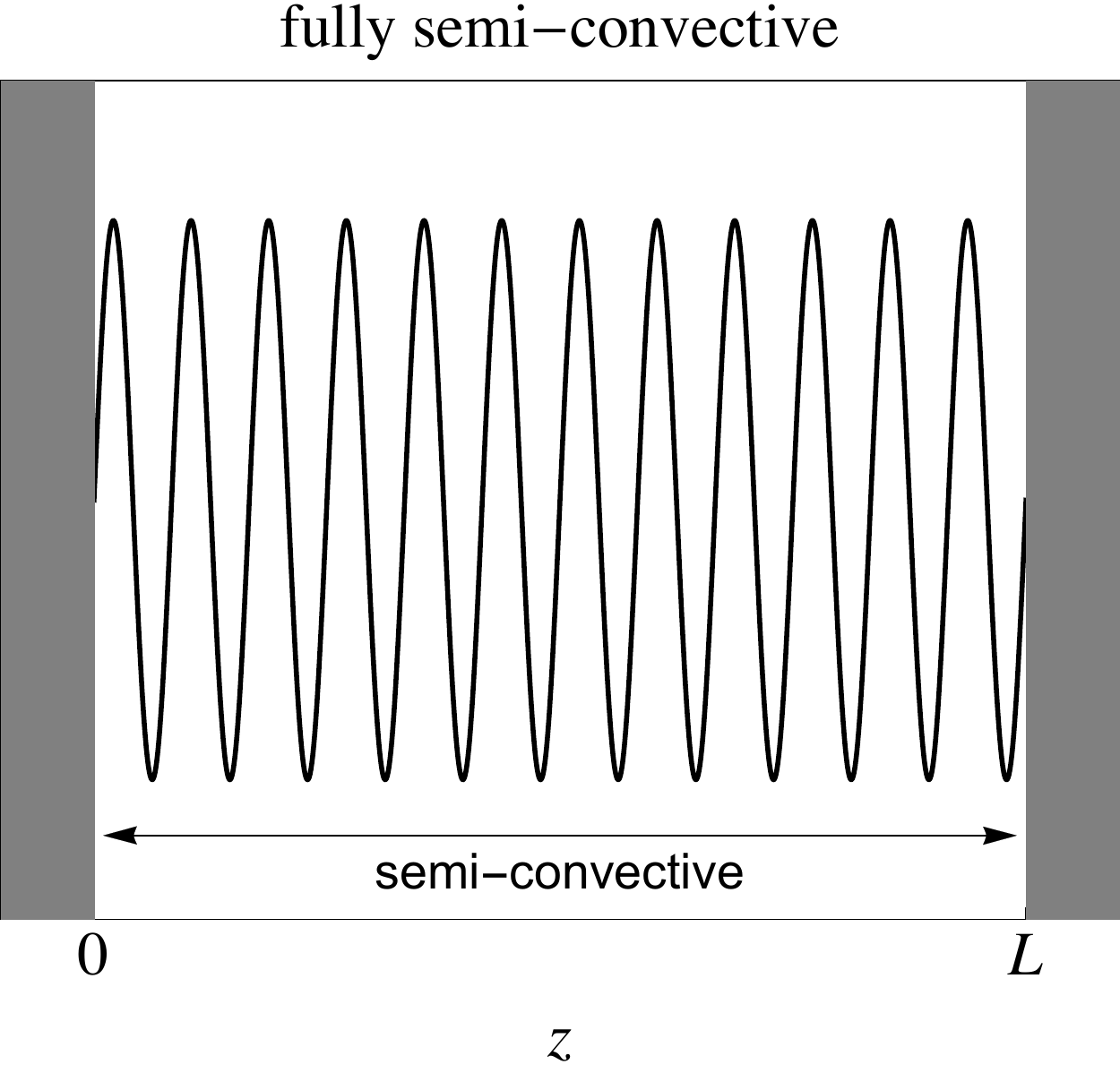}}
\subfigure{\includegraphics[width=.48\textwidth]{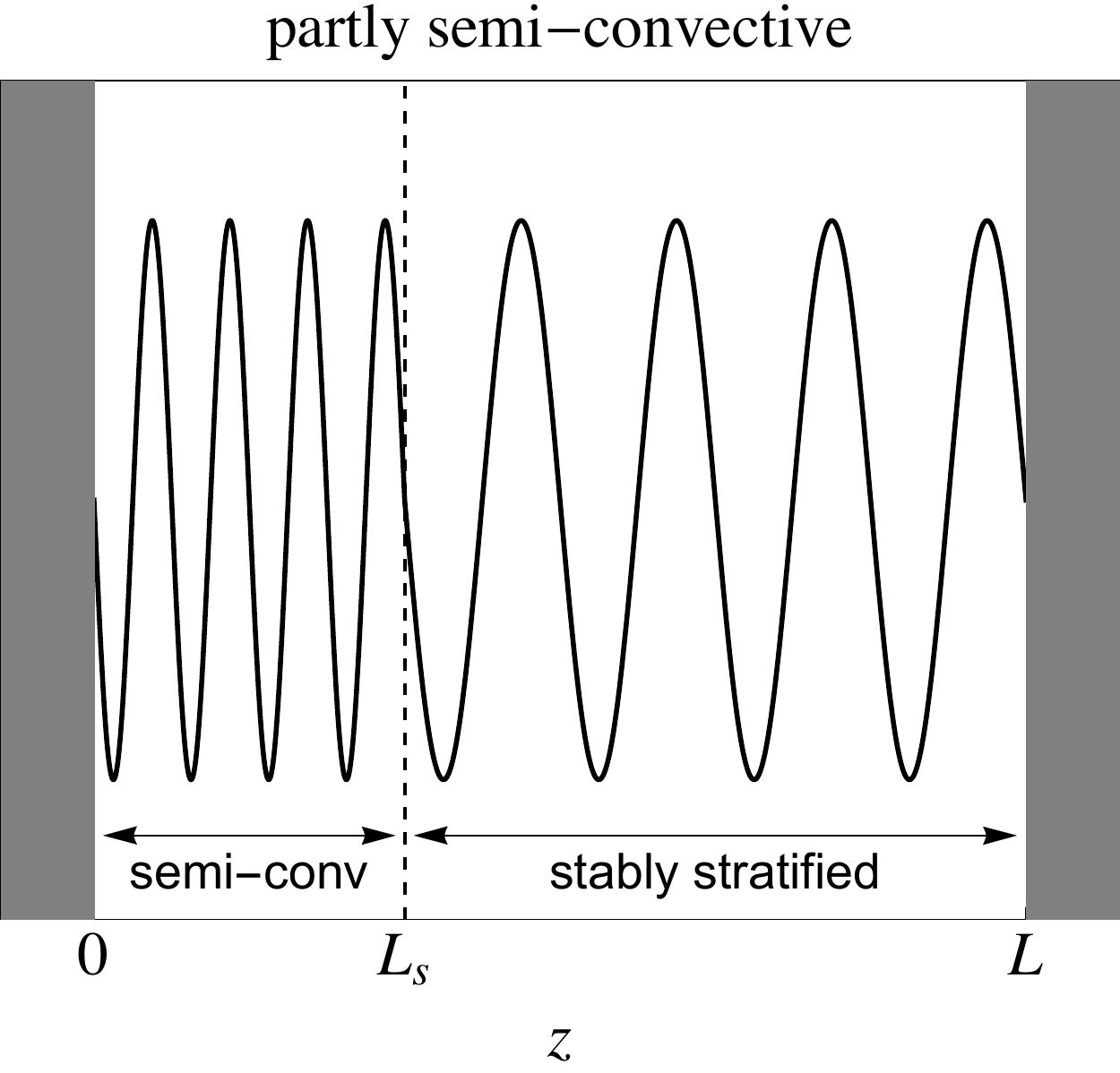}}
\caption{Schematic of a g-mode of a given frequency trapped between bounding walls (gray) in a cavity of size $L$ (in the $z$-dimension). Left panel: the g-mode cavity consists of a single zone of layered semi-convection with constant $\bar{N}$. Right panel: the g-mode cavity contains a zone of layered semi-convection of size $L_s$ adjacent to a stably stratified zone of size $L-L_s$. The value of the \BV frequency in the stably stratified zone is also $\bar{N}$. However, due to a modified (stiffer) dispersion relation in the semi-convective zone, $k_z$ is greater there than in the stably stratified zone for a given $\omega$.}
\label{cavity_fig}
\end{figure*}

In what follows, we consider a cavity of fixed size with reflecting walls in a plane parallel medium that has constant gravity and \BV frequency. We then ask how the constant period spacing relation is affected by layered semi-convection within the cavity? For our first example, we consider a g-mode cavity that has layered semi-convection everywhere (left panel of Fig.\ \ref{cavity_fig}). This is what one might expect for a giant planet with semi-convection throughout the interior (or at least throughout the stably stratified region). For our second example, we consider a cavity that is partly stably-stratified and partly semi-convective (right panel of Fig.\ \ref{cavity_fig}). This is what one would expect for e.g. a star undergoing helium burning with a convective core. For simplicity, we do not consider rotation, but it is possible to include it in the analysis. 

For a fully semi-convective cavity with constant $\bar{N}$, $k_z$ is constant for a given value of the period $P \equiv 2 \pi/\omega$. Because the cavity has reflecting walls, an integer number of wavelengths must fit inside the cavity. Consequently, 
\begin{align}
\label{kzn}
k_z = 2\pi n/L,
\end{align} 
where $n \ge 1$ is an integer and $L$ is the size of the cavity. As we have shown in \S \ref{comparison_sec}, in the limit $d \rightarrow 0$, we recover (to zeroth order in $d$) the dispersion relation for a continuously stratified medium with \BV frequency $\bar{N}$. Written in terms of $P$ and $n$, the dispersion relation for a continuosly stably stratified medium is
\begin{align}
\label{Pdisrel}
P_0 \approx \frac{(2 \pi)^2 n}{\bar{N}k_\perp L}.
\end{align}
The period spacing relation can be obtained by differentiating equation (\ref{Pdisrel}) with respect to $n$, which gives
\begin{align}
\label{Pspacing}
\Delta P_0 = \frac{(2 \pi)^2}{\bar{N}k_\perp L}.
\end{align}
This is the period change in going from $n$ to $n+1$ nodes in the $z$ dimension. Note that $\Delta P_0$ is a constant independent of $n$ (i.e. the mode frequency).

From equation (\ref{deltaomega}), it is clear that for a semi-convective staircase there is a second order correction to the frequency, and hence to the period, in both $k_\perp d$ and $k_z d$ compared to a continuously stably-stratified medium. Astrophysically relevant g-modes typically have spherical harmonic degree $l=1$ or $l=2$. Since $k_\perp \sim l/r_*$ in our setup, $k_\perp d \sim l d /r_*$. Because a semi-convective staircase extends over a region of the star that is a small fraction of the stellar radius, the period spacing correction in the semi-convective staircase due to $k_\perp d$ is too small to be observable. However, $k_z$ is larger than $k_\perp$ by a factor of $(n/l) (r_*/L)$, and for $n \gg 1$ the period spacing correction due to semi-convection {\it is} potentially observable. We work out that correction below: first for a g-mode cavity that has layered semi-convection throughout (left panel of Fig.\ \ref{cavity_fig}) and subsequently for a cavity that is partly semi-convective (right panel of Fig.\ \ref{cavity_fig}). 

For layered semi-convection, we can use equation (\ref{deltaomega}) to write
\begin{align}
P \approx \frac{2 \pi k_z}{\bar{N}k_\perp }\left(1 - \frac{1}{24}(k_z d)^2\right),
\end{align}
which is valid to second order in $d$ in the limit $k_z d \ll 1$. Equation (\ref{kzn}) still applies, because by assumption the properties of the medium are independent of $z$ within the cavity, so we can write
\begin{align}
\label{Pneq}
P \approx \frac{(2 \pi)^2 n}{\bar{N}k_\perp L}\left(1 - \frac{1}{24}\left(\frac{2 \pi n d}{L}\right)^2 \right).
\end{align} 
Differentiating equation (\ref{Pneq}) with respect to $n$, the period spacing relation with the correction for layered semi-convection is
\begin{align}
\label{DeltaPfullylayered}
\Delta P = \Delta P_0 \left(1 - \frac{1}{2}\left(\frac{\pi n d}{L}\right)^2 \right).
\end{align}

From equation (\ref{DeltaPfullylayered}), we see that the g-mode period spacing is smaller for the staircase than for a continuously stratified medium. We also see that the g-mode period spacing decreases with increasing $n$, or alternatively decreasing frequency. Moreover, the correction due to layered semi-convection becomes more pronounced as $n$ increases, which is the opposite of e.g. the correction due to a finite scale height, which becomes smaller with increasing $n$. Although our approximation (\ref{DeltaPfullylayered}) is not valid for $2\pi nd/ L \sim 1$ (or equivalently $k_z d\sim 1$), we can understand what happens in this limit by looking at the right panel of Fig \ref{compare_fig}. From the figure, we see that at the cutoff frequency, $\omega = \omega_c$, the quantity $d\omega/d k_z = 0$. This implies that $d P/dn = 0$ at $\omega = \omega_c$, so that the period spacing tends to zero as we approach the cutoff frequency. Below the cutoff frequency waves do not propagate in the layered semi-convective medium.

Next we consider the period spacing relation for a g-mode cavity which contains a region that is stably stratified adjacent to a region undergoing layered semi-convection (right panel of Fig. \ref{cavity_fig}). We reiterate that this is common in post main sequence stellar interiors. One can show that for $k_z d \ll 1$, the analog of equation (\ref{DeltaPfullylayered}) for a g-mode cavity that is partly semi-convective is 
\begin{align}
\label{DeltaPpartlylayered}
\Delta P = \Delta P_0 \left(1 - \frac{1}{2}\left(\frac{\pi n d}{L}\right)^2 \frac{L_s}{L} \right),
\end{align}
where $L_s/L$ is the ratio of the size of the semi-convective region to the total size of the g-mode cavity. Once again, we we see that the period spacing is smaller for the model with layered semi-convection and $\Delta P$ decreases for larger $n$, i.e. lower frequency g-modes. 

Comparing equations (\ref{DeltaPfullylayered}) and (\ref{DeltaPpartlylayered}), one sees that the magnitude of the correction to the period spacing relation is proportional to $L_s/L$, so the correction becomes larger as semi-convection fills a fractionally larger volume of the cavity. Another major difference compared to a fully semi-convective cavity is that modes with frequencies $\omega < \omega_c$ can propagate within the cavity, but only in the stably stratified region, not in the region of layered semi-convection (see equation (\ref{cutoff_freq})). Thus, there is an approximately constant period spacing $\Delta P \approx (2 \pi)^2/\bar{N}k_\perp L$ for $\omega \gg \omega_c$ and a {\it different} approximately constant period spacing $\Delta P \approx (2 \pi)^2/\bar{N}k_\perp (L-L_s)$ for $\omega \ll \omega_c$. The transition between the two period spacings occurs at frequencies $\omega \sim \omega_c$ (equation (\ref{cutoff_freq})).

\section{Discussion}
\label{discussion}

We have analyzed the dispersion relation of g-modes (buoyancy waves) propagating through a plane-parallel fluid which has the structure of a layered semi-convective staircase. We assumed that the size of the stairs in the semi-convective staircase was small compared to the scale height. The fluid was assumed to be perfectly convective inside of a stair (vanishing entropy gradient and \BV frequency), and we imposed a small (relative to the background density), discontinuous jump in the density at the interfaces between stairs (see Fig.\ \ref{density_schematic}). Although this simplified model abstracted away the rich, double-diffusive physics leading to semi-convection, its simplicity allowed us to derive analytical dispersion relations for g-modes in a semi-convective staircase. Such analytical results are particularly relevant given the rich asteroseismic data emerging from e.g.\ the {\it Kepler} mission and can serve as guide posts for understanding the effect of a region of layered semi-convection on the propagation of g-modes in a star or planet.

Using a matrix transfer formalism \citep{Molinari08}, we were able to find analytical solutions for the dispersion relation of g-modes in layered semi-convection under either the assumption of an infinitely extended staircase (\S \ref{disrel_periodic},\ref{disrel_infinite}) or a staircase of finite vertical extent embedded in a fully-convective medium (\S \ref{disrel_finite}). We were also able to quantify the effect of rotation on the dispersion relation of g-modes and explicitly determine how the properties transition from pure buoyancy waves to inertial waves as the rotation rate increases (\S \ref{rotationsec}). Note that the latter are unaffected by the presence of a staircase.

The dispersion relation for g-modes in layered semi-convection is stiffer -- i.e. the frequency is higher for a given wavelength -- than for internal gravity waves in a continuously stratified medium. Modes with wavelengths that are long compared to the step size behave like gravity modes with a leading order fractional correction to the frequency of $\mathcal{O}[(d/\lambda)^2]$, where $\lambda$ is the wavelength and $d$ is the distance between layers in the staircase (see equation (\ref{deltaomega})). On the other hand, frequencies of g-modes with wavelengths comparable to the step size are strongly affected by the discreteness of the density jumps in the staircase. One of the most dramatic manifestations of this difference is a lower cutoff frequency for the staircase, which does not exist for a continuously stratified medium. The effect of this cutoff frequency is to prevent the propagation of modes with vertical wavelengths shorter than the step size in the staircase. 

The stiffer dispersion relation and the presence of a cutoff frequency both have implications for detecting layered semi-convection using asteroseismology (see \S \ref{asteroseismology_sec}). In particular, layered semi-convection decreases the period spacing of g-modes relative to that expected for a continuously stratified medium (\S \ref{asteroseismology_sec} and equations \ref{DeltaPfullylayered} and \ref{DeltaPpartlylayered}).  Moreover, the decrease in the period spacing is frequency dependent, becoming larger for lower frequency (shorter wavelength) g-modes that are more affected by the discrete density jumps in the semi-convective staircase. 

The change in g-mode period spacing due to layered semi-convection we predict in equations (\ref{DeltaPfullylayered},\ref{DeltaPpartlylayered}) can be differentiated from other physical effects.  Non-WKB corrections to the period spacing of g-modes become smaller for lower frequency waves, in contrast to the semi-convective corrections which become larger for lower frequency waves.  In addition, mode trapping (which can also be created by composition gradients but is due to non-WKB effects on the mode frequencies) creates localized ``dips" in the period spacing (see, e.g., \citet{Papics2015}). Rapid rotation often creates a period spacing that changes nearly linearly with radial order $n$ \citep{Bouabid}, with an opposite slope for prograde and retrograde modes.

Several well-measured period spacings in pulsating stars with convective cores (which may contain semi-convective regions at the core boundary) have been presented in \citet{Papics2014,Papics2015,Saio,vanReeth}. None of them clearly exhibit a quadratic dependence on $n$, or a transition from a larger period spacing to a smaller one. This may provide evidence that semi-convective regions do not occupy large portions of low-mass stellar interiors, as found by \citep{MooreGaraud}.

Although we have focused on g-modes, one may also wonder how a semi-convective staircase affects the propagation of p-modes. It is more difficult to obtain analytical results in this case, but it is still possible to apply the matrix transfer formalism used for g-modes. In so doing, we generically find that the leading order fractional correction to p-mode frequencies is $\mathcal{O}[(d/H)^2]$, which is much less than for g-modes when the wavelength is less than the scale height (i.e. $\lambda \ll H$). Moreover, unlike for g-modes, there is no lower cutoff frequency introduced by the finite size of the steps (though there is still a lower cutoff at $\bar{N}$, the analog of the \BV frequency introduced in equation (\ref{barN})). Both of these reasons make g-modes much better candidates than p-modes for identifying regions of semi-convection in asteroseismic power spectra.   

An effect which we have not included in our analysis, but which is likely to be important in astrophysical semi-convection is the forcing of modes by convective eddies. In particular, the presence of convection within a stair should preferentially excite modes with frequencies that are of order the eddy turnover frequency. Indeed, excitation of modes has been observed in simulations of layered semi-convection (e.g.\ \citet{Wood13}) and measuring the frequencies of such modes in simulations would be a useful check on the assumptions used to derive our analytical results.

\section*{Acknowledgments}
We thank Pascale Garaud for fruitful discussions on stellar and planetary oscillations and semi-convection. This work was supported in part by the Theoretical Astrophysics Center at UCB, by a Simons Investigator award from the Simons Foundation to EQ, and by the David and Lucile Packard Foundation. JF acknowledges partial support from NSF under grant no. AST-1205732 and through a Lee DuBridge Fellowship at Caltech

\bibliography{../bibliography/modesdd}

\appendix
\section{Jump Conditions at the Interfaces in the Presence of Rotation}
We derive jump conditions at the interfaces between steps in the presence of a constant rotation rate. We start by writing down the linearized fluid equations in the Boussinesq approximation in the corotating frame. The direction of stratification is along the $z$ axis, and without loss of generality, we take $\bfOmega = \Omega_x \xhat + \Omega_z \zhat$. The primes denote differentiation with respect to $z$, and we have Fourier transformed in $x$, $y$, and $t$:
\begin{align}
\label{rot_equations}
0 &= i k_x \dd u + i k_y \dd v + \dd w' \\
i \omega \dd u &= \frac{i k_x \dd P}{\rho} - 2\Omega_z \dd v \nn \\ 
i \omega \dd v &=  \frac{i k_y \dd P}{\rho} - 2\Omega_x \dd w +2\Omega_z \dd u \nn \\ 
i \omega \dd w &= \frac{\dd P'}{\rho} + 2\Omega_x \dd v + g \frac{\delta \rho}{\rho} \nn \\ 
i \omega \frac{\gamma \dd \rho}{\rho} &= -\dd w \frac{d \ln P \rho^{-\gamma}}{dz} \nn
\end{align}
Note that as in \S \ref{rotationsec}, we have ignored the centrifugal force in the equations of motion, which assumes the rotation rate is far below breakup. 

\subsection{Rotation Axis Along $z$}
\label{aligned_app}
When the spin axis is along the direction of stratification ($\Omega_x = 0$) we can set $k_y = 0$ without loss of generality. The set of equations (\ref{rot_equations}) can be combined into a pair of first order partial differential equations,
\begin{align}
\label{two_equations}
\left(1 - \frac{\bar{N}^2}{\omega^2} \right) \dd w &= \frac{\dd P'}{i \omega \rho} \\
\dd P &= \frac{i \omega \rho}{k_x^2} \left(1 - \frac{4 \Omega_z^2}{\omega^2} \right) \dd w' , \nn
\end{align}
where $N(z)$ is defined via equation (\ref{Bruntdef}) and the primes denote differentiation with respect to $z$. The set of equations (\ref{two_equations}) can be combined into a single second order partial differential equation for $\dd w$:
\begin{align}
\label{master_eq}
\rho k_x^2 \left(\omega^2 - \bar{N}^2\right) \dd w = \left(\omega^2 - 4 \Omega_z^2 \right)\left(\rho \dd w' \right)'.
\end{align}

Because equation(\ref{master_eq}) is a second order differential equation, we need two jump conditions on $\delta w$ across the interface. Using the subscript ``0" for the fluid above the interface and the subscript ``1" for the fluid below the interface, the first of these conditions is $\dd w_0 = \dd w_1$, which expresses the requirement that the upper and lower fluids stay in contact across the interface. 

To obtain the second jump condition, we integrate equation (\ref{master_eq}) across the interface to obtain
\begin{align}
\label{jump2}
-g \frac{\Delta \rho}{\rho} \dd w = \frac{\omega^2-4\Omega_z^2}{k_x^2}\left(\delta w_0' - \delta w_1' \right).
\end{align}
Here, we have taken $\Delta \rho = \rho_1-\rho_0$ and assumed that $\Delta \rho/\rho \ll 1$. The $g \Delta \rho/\rho$ term comes from integrating $N^2$ across the interface. 

The interface conditions (\ref{BC_AB}) are then modified to read
\begin{align}
A_{n+1}e^{\bar{k}d}-B_{n+1}e^{-\bar{k}d} &= A_n-B_n  \\ \nn
A_{n+1}e^{\bar{k}d} + B_{n+1}e^{-\bar{k}d} &= A_n + B_n +\frac{g \bar{k}}{\omega^2}\frac{\Delta \rho}{\rho}(A_n-B_n),
\end{align}
where $\bar{k}$ is given by equation (\ref{k_aligned}). The dispersion relation (\ref{exact_disrel_aligned}) follows directly.

\subsection{Rotation Axis Perpendicular to $z$}
\label{misaligned_app}
When the spin axis is perpendicular to the direction of stratification ($\Omega_z = 0$), the set of equations (\ref{rot_equations}) can be combined into the following pair of first order partial differential equations
\begin{align}
\label{two_equations_perp}
\left(1 - \frac{N^2}{\omega^2} - \frac{4 \Omega_x^2}{\omega^2}\right) \dd w &= \frac{\dd P'}{i \omega \rho} + \frac{2 \Omega_x k_y}{i \omega^2 \rho} \dd P \\
 \dd P &= - \frac{i \rho}{k_x^2 + k_y^2} \left( 2\Omega_x k_y \dd w - \omega \dd w' \right). \nn
\end{align}
The set of equations (\ref{two_equations_perp}) can further be combined into a single second order partial differential equation for $\dd w$:
\begin{multline}
\label{master_eq_perp}
\rho \left(\omega^2 - N^2 - 4 \Omega_x^2 \right) \dd w = \\ \frac{\left(-4 k_y^2 \Omega_x^2 \rho -2 k_y \Omega_x \omega \rho'\right) \dd w + \omega^2 \left(\rho \dd w' \right)'}{k_x^2 + k_y^2}.
\end{multline}

To obtain the jump condition for $\dd w'$, we integrate equation (\ref{master_eq_perp}) across the interface to obtain
\begin{align}
\label{jump2_perp}
-\left(\frac{2k_y \Omega_x \omega}{k_x^2+k_y^2} + g \right) \frac{\Delta \rho}{\rho} \dd w = \frac{\omega^2}{k_x^2 + k_y^2}\left(\delta w_0' - \delta w_1' \right).
\end{align}

We can simplify equation (\ref{jump2_perp}) further by showing that the first term in parentheses on the left hand side is much smaller than $g$, under the Boussinesq assumption. Note that $\omega k_y/(k_x^2+k_y^2) < \omega/\sqrt{k_x^2+k_y^2} \ll c_s$, where $c_s$ is the sound speed. Since the scale height is $H \sim c_s^2/g$, we can write:
\begin{align}
\frac{k_y \Omega \omega}{k_x^2+k_y^2} \ll \Omega c_s  \sim \Omega \sqrt{gH} \ll g.
\end{align}
The last inequality follows from the fact that $\Omega^2 H < \Omega^2 R \ll g$, which comes from our assumption that the centrifugal force is dynamically unimportant. Thus, equation (\ref{jump2_perp}) simplifies to
\begin{align}
\label{jump2_simple}
- g \frac{\Delta \rho}{\rho} \dd w = \frac{\omega^2}{k_x^2 + k_y^2}\left(\dd w_0' - \dd w_1' \right).
\end{align}

The interface conditions (\ref{BC_AB}) are then modified to read
\begin{align}
A_{n+1}e^{\bar{k}d}-B_{n+1}e^{-\bar{k}d} &= A_n-B_n  \\ \nn
A_{n+1}e^{\bar{k}d} + B_{n+1}e^{-\bar{k}d} &= A_n + B_n +\frac{g \left(k_x^2 + k_y^2\right)}{\bar{k} \omega^2}\frac{\Delta \rho}{\rho}(A_n-B_n),
\end{align}
where $\bar{k}$ is given by equation (\ref{k_misaligned}). The dispersion relation (\ref{exact_disrel_misaligned}) follows directly.

\end{document}